\documentclass[a4paper,11pt]{article}
\usepackage{pos}

\title{Energy-momentum tensor in the 2D Ising CFT in full modular space}
\ShortTitle{Energy-momentum tensor in the 2D Ising CFT}

\author[a]{Richard C.~Brower}
\author[b]{George T.~Fleming}
\author*[a]{Nobuyuki Matsumoto}
\author[a]{Rohan Misra}

\affiliation[a]{Boston University,
  Boston, MA 02215, USA}
\affiliation[b]{Fermi National Accelerator Laboratory,
Batavia, Illinois, 60510, USA}

\emailAdd{nmatsum@bu.edu}

\abstract{ 
A set of lattice operators for the energy-momentum (EM) tensor in the Ising CFT 
is derived 
in the spin variables.
Our expression works under 
arbitrary affine transformation
both on triangular and hexagonal 
lattices (where the former includes 
the rectangular lattices).
The correctness of the operators
is numerically confirmed
in Monte Carlo calculations
by comparing the results with the conformal Ward identity,
including the
operator 
normalization.
In the derivation of the EM tensor,
a staggered structure 
of the affine-transformed 
hexagonal lattice is analyzed,
which shows a peculiar shift from
the circumcenter dual lattice
and appears as a mixing angle between
the holomorphic part $T(z)$ and
the antiholomorphic part $\tilde T(\bar z)$.
The details of this contribution
will appear in a subsequent 
paper.}

\FullConference{The 41st International Symposium on Lattice Field Theory (LATTICE2024)\\
 28 July - 3 August 2024\\
Liverpool, UK\\}

\begin{document}
\maketitle

\section{Introduction}

To expand 
the 
non-perturbative understanding 
of quantum field theory,
the Quantum Finite Elements (QFE) project \cite{Brower:2016moq}
aims to 
enable lattice calculation
on curved spacetime. 
Such an extension
enables us to,
for example, 
use radial quantization 
for conformal field theories 
(CFTs) in 
the spacetime
with dimensions $D>2$
\cite{Fubini:1972mf,Cardy:1985lth,Brower:2012vg}.
As a step towards 
its rigorous formulation,
this work considers the 
energy-momentum (EM) tensor
in the 2D CFT on the torus
with generic modulus $\tau$,
which is a key physical object
that generates diffeomorphism. 
We derive a lattice
expression of the EM tensor
in both fermionic and 
spin variables.
The expression is 
confirmed in direct Monte Carlo calculation
by checking the 
conformal Ward identities~\cite{Eguchi:1986sb,DiFrancesco:1987ez}.
Although 
we are not the first people
who address the EM tensor
in the two-dimensional Ising spin model~\cite{Kadanoff:1970kz},
we derive a complete expression
including normalization
that works under
an arbitrary 
affine transformation.
To our knowledge,
our work is the first work
that verifies the 
analytic formulas
for the Ising CFT 
involving the EM tensor
numerically with 
the spin variables.
The details of the content
will be given in a
subsequent paper \cite{NM2025}.

\section{Affine-transformed lattice}

\begin{figure}[htb]
  \centering
  \includegraphics[width=9.5cm]{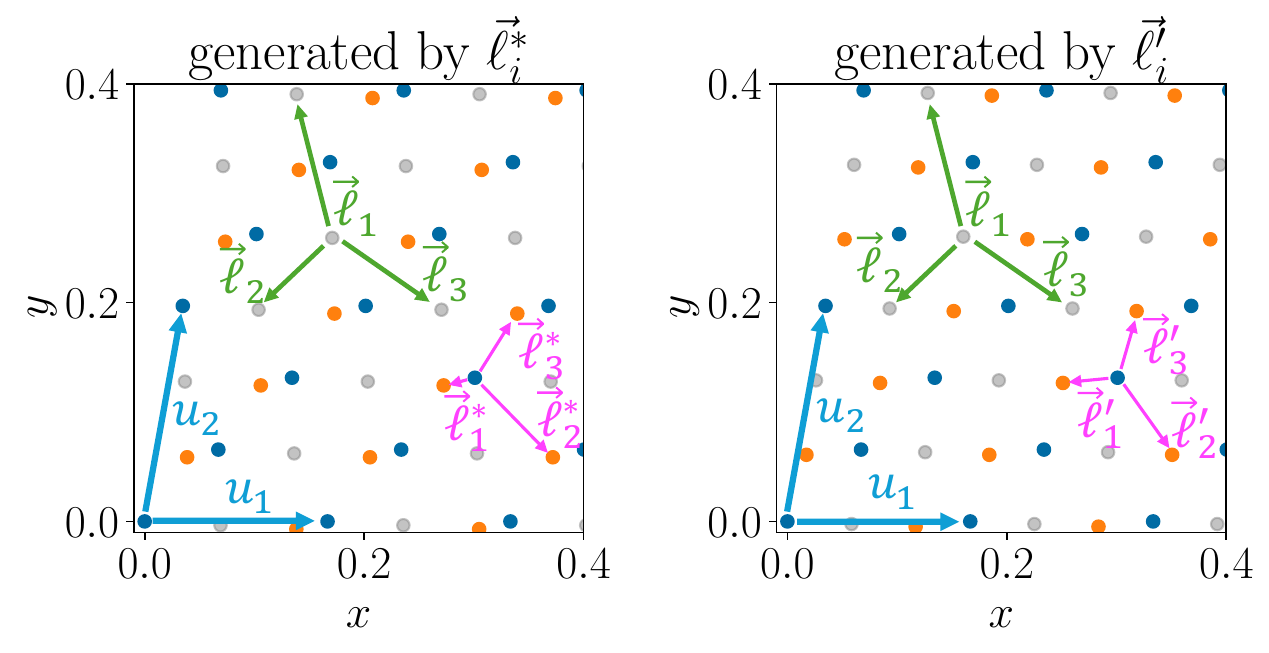}
  \caption{Gray dots represent the affine-transformed triangular lattice. 
  The blue and orange points
  represent its dual lattices,
  where the colors 
  distinguish the even (blue) and odd (orange) sites.
  The left panel shows
  the circumcenter dual lattice,
  generated by $\vec \ell_i^*$,
  while the right panel
  shows the lattice generated by
  $\vec \ell_i'$, on which 
  the fields reside
  (see Sec.~\ref{sec:staggered}).
  }
  \label{fig:lattices}
\end{figure}

We consider a generic 
affine-transformed 
triangular lattice
as shown in Fig.~\ref{fig:lattices}.
We write 
the Cartesian coordinates 
as $(x_\mu) \equiv (x,y)$
and the three lattice translation
vectors as
$\vec \ell_k$ $(k=1,2,3)$.
The circumcenter dual 
of the triangular lattice
is a hexagonal lattice,
and we write the 
three lattice translation vectors
from even to odd sites as $\vec \ell_k^*$.
We further define the unit
vectors on the hexagonal lattice:
  $\vec e_k \equiv \vec \ell^*_k/\vert \vec \ell^*_k \vert
\equiv ( \cos \alpha_k, \sin \alpha_k )^T.$
Depending on the context,
we use an alternative notation
such as $\vec \ell^*_{nm}$
for the lattice translation
vector from the site $n$ to $m$.

We impose periodicity of the torus 
by the identification:
\begin{align}
  0 \sim L_1 \vec u_1 \sim L_2 \vec u_2
  \quad
  (L_1, L_2 \in \mathbb{Z}),
\end{align}
where
$\vec u_1 \equiv -2\vec \ell_1^* + \vec \ell_2^* + \vec \ell_3^*$
and
$\vec u_2 \equiv -\vec \ell_1^* - \vec \ell_2^* + 2\vec \ell_3^*$. 
In this work, 
we set $L_1=L_2 \equiv L$
for simplicity.
As we consider a conformal theory on the torus,
the scale of the system
can be conventionally
fixed by setting 
one of the lattice dimension
to unity:
$L |\vec u_1| = 1$.
With the 
standard identification
between the two-dimensional vector and 
the complex variable:
$\mathbb{R}^2\simeq \mathbb{C}$,
$\vec u = (u_x, u_y)^T \leftrightarrow u \equiv u_x +i u_y$,
the modulus of the torus
can be expressed as $\tau \equiv L u_2$.

\section{Partition functions}

As is well known,
the 2D Ising model
has a fermionic description
with the Majorana fermion \cite{PhysRev.65.117,RevModPhys.36.856,Itzykson:1982ed}.
For later discussion,
we consider 
the Wilson-Majorana 
fermion action \cite{Wolff:2020oky,Brower:2022cwv}
on the hexagonal lattice
in a generalized setup
with arbitrary couplings:
\begin{align}
  S_W
  &=
    \frac{1}{2}\sum_{n}
    (1+\Delta m_n)
    \bar \psi_{n} \psi_{n}
    -
    \sum_{ \langle n, m \rangle}
    \kappa_{nm}
    \,
    \bar \psi_{n}
    P(\vec e_{nm})
    \psi_{m},
    \label{eq:Slat}
\end{align}
where $\psi \equiv (\psi_1, \psi_2)^T$
is the two-component
real Grassmann 
variable,
$\langle n, m \rangle$
denotes the nearest neighbor pair,
and
$P(\vec e)$ is the Wilson projector:
\begin{align}
  P(\vec e)
  \equiv
  \frac{1}{2}(1 - \vec e \cdot \vec \sigma).
\end{align}
We use the Pauli matrices
$\vec\sigma \equiv (\sigma_1, \sigma_2)$
as the gamma matrices throughout the paper.
The partition function 
for the lattice fermion is then:
\begin{align}
  Z_W^{\rm hex}
  \equiv
  \int {\cal D}\psi\,
  e^{-S_W},
    \quad
    {\cal D}\psi
  \equiv
  \prod_n
  d \psi^1_{n}
  d \psi^2_{n}.
  \label{eq:ZMlat}
\end{align}

Our main focus is the
Ising spin system
on triangular and hexagonal lattices,
whose partition functions are:
\begin{align}
  Z_{I}^{\rm tri}
  & \equiv
  \sum_{\{s_k=\pm 1\}}
  \exp\Big[
  \sum_{\langle k,\ell \rangle}
  L_{k\ell}
  s_k s_\ell
  \Big],
  \label{eq:ZItri}
  \\
  Z_{I}^{\rm hex}
  & \equiv
  \sum_{\{\mu_n = \pm 1\}}
  \exp\Big[
  \sum_{\langle n,m \rangle}
  K_{nm}
  \mu_n \mu_m
  \Big].
  \label{eq:ZIhex}
\end{align}
The partition functions~\eqref{eq:ZMlat},
\eqref{eq:ZItri} and \eqref{eq:ZIhex}
can be related to each other
by loop expansions \cite{10.1063/1.524404,Wolff:2020oky,Brower:2022cwv}
even under various boundary conditions
on the torus.
For the discussion below, 
it suffices to quote the following identities:
\begin{align}
  \frac{1}{
  2^{N_{\rm site}^{\rm hex}}
  \prod_{\langle n,m \rangle} \cosh K_{nm}
  } Z_I^{\rm hex}(P,P)
  &=
    \frac{1}{\prod_{n} (1+\Delta m_n) }
    \Big[
    -
    Z_W^{\rm hex}(P,P)
    +
    \sum_{\varepsilon \neq (P,P)}
    Z_W^{\rm hex}(\varepsilon)
    \Big],
    \label{eq:hex_hex}\\
  \frac{1}{
  2 \prod_{\langle i,j \rangle}
  \exp L_{ij}
  }
  Z_I^{\rm tri}(P,P)
  &=
    \frac{1}{\prod_{n}(1+\Delta m_n)}
    \Big[
    +
    Z_W^{\rm hex}(P,P)
    +
    \sum_{\varepsilon \neq (P,P)}
    Z_W^{\rm hex}(\varepsilon)
    \Big].
    \label{eq:tri_hex}
\end{align}
Here, the couplings are
identified as:
\begin{align}
  \frac{\kappa_{nm}}{\sqrt{(1+\Delta m_n)(1+\Delta m_m)}}
  \frac{
  \cos \Delta \alpha_{n m n'}
  \cos \Delta \alpha_{n m n'' }}{
  \cos \Delta \alpha_{n' m n''} }
  =
  \tanh K_{nm}
  =
  e^{-2L_{ij}},
\end{align}
where the angle differences 
$\Delta \alpha_{n m n'} \equiv \alpha_{nm} - \alpha_{m n'}$
are taken among the 
three neighboring sites from $m$: $n$, $n'$, and $n''$.
In eqs.~\eqref{eq:hex_hex} and
\eqref{eq:tri_hex},
the boundary conditions
are indicated by 
$\varepsilon = \textrm{(P,P)},
\textrm{(A,P)},
\textrm{(A,A)},
\textrm{(P,A)}$,
where P stands for periodic
and A for antiperiodic.
Different signs for the
$\textrm{(P,P)}$ sector 
correspond to the sign ambiguity
in defining the chirality operator
in the Ising CFT~\cite{Seiberg:1986by,NM2025}.

\section{Staggered lattice structure in the affine-transformed hexagonal lattice}
\label{sec:staggered}

In Ref.~\cite{Brower:2022cwv},
it was shown that
the continuum limit 
can be taken 
under nontrivial affine transformation
with the couplings:
\begin{align}
  \Delta m_n = 0,
  \quad
  \kappa_{n, n+\hat k}
  =
  \frac{2 \vert \vec \ell_k \vert}
  {
  \sum_{k'} \vert \vec \ell_{k'} \vert
  }.
  \label{eq:crit_sol_affine}
\end{align}
Although this is correct,
it turns out that
the bipartite hexagonal lattice 
shows a peculiar staggered structure
under nontrivial affine transformation.
While the directional information
of the circumcenter dual lattice,
$e_k$, 
appears in the lattice action~\eqref{eq:Slat},
the fermionic variables turn out to
reside on a different lattice
generically.
This fine structure of the 
hexagonal lattice
is relevant in 
deriving the lattice expression
of the EM tensor 
as it involves 
a derivative in 
fermionic variables.

To see this, 
we consider the 
lattice equation of motion (EOM), $E_{\psi_n} = 0$,
where
\begin{align}
  E_{\psi_n}
  &=
  \psi_n
  -
  \frac{1}{2} 
  \sum_k
  \kappa_k
  \big(
  1 - \vec e_k \cdot \vec \sigma
  \big)
  \psi_{n+\hat k}
  \quad (n : \textrm{even}), \\
E_{\psi_n}
  &=
  \psi_n
  -
  \frac{1}{2} 
  \sum_k
  \kappa_k
  \big(
  1 + \vec e_k \cdot \vec \sigma
  \big)
  \psi_{n-\hat k}
  \quad(n : \textrm{odd}).
\end{align}
The action is quadratic, 
and the infrared property of the 
Wilson-Dirac operator
can be analyzed by the derivative
expansion:
\begin{align}
  \psi_{n+\hat k} \simeq \psi_n + \ell'_{k\mu} \partial_\mu \psi_n,
  \label{eq:ell_expansion_psi}
\end{align}
where the translation vectors
$\vec \ell'_k$ 
are arbitrary for now
and to be determined below.
For even sites:
\begin{align}
  E_{\psi_n}
  &\simeq
    \Big(
    1-\frac{1}{2}\sum_{k}
    \kappa_k
    \Big)
    \psi_n
    +
    \frac{1}{2}
    \Big( \sum_k \kappa_k e_{k\mu} \Big)
    \sigma_\mu \psi_n
    \nonumber\\
  &~~~~~~~~~~~
    -
    \frac{1}{2}
    \Big( \sum_k \kappa_k \ell'_{k\mu} \Big)
    \partial_\mu \psi_n
    +
    \frac{1}{2}
    \Big( \sum_k \kappa_k e_{k\mu} \ell'_{k\nu} \Big)
    \sigma_\mu \partial_\nu \psi_n,
\end{align}
while for odd sites
the signs of the second and third 
terms flip.
The lattice 
EOM operators
should approach the
continuum expression:
\begin{align}
    E_{\psi}^{\rm cont}(x)
    \equiv
    \sigma_\mu \partial_\mu \psi(x)
\end{align}
up to proportionality constant.
Under the solution~\eqref{eq:crit_sol_affine},
this requirement reduces 
to the condition for 
the vectors $\vec{\ell}'_{k}$:
\begin{align}
  \sum_k \kappa_k \ell'_{k\mu} \equiv 0 ,
  \quad
  \sum_k \kappa_k e_{k \mu} \ell'_{k\nu} \equiv
  2 a \delta_{\mu\nu},
  \label{eq:cond_ell_tilde}
\end{align}
where a scaling constant $a$
can be set by demanding 
again the
lattice dimension in the $x$
direction to be unity.
We have six conditions for 
six independent real variables
(where an overall scaling is fixed by $a$),
and the condition is sufficient to determine
$\vec{\ell}'_{k}$.

As emphasized above, 
the solution 
$\vec{\ell}'_{k}$
to eq.~\eqref{eq:cond_ell_tilde}
does not agree with 
$\vec{\ell}^*_{k}$
under 
a generic affine transformation.
In Fig.~\ref{fig:lattices},
the two lattices are compared
for the case $\tau=1.2 e^{4i\pi/9}$. 
We see that 
the lattice generated by 
$\vec{\ell}'_{k}$ has more
regular appearance 
than that generated by 
$\vec{\ell}^*_{k}$.

\section{Lattice operators}

We are now ready to derive
the lattice EM tensor
in spin variables 
by applying 
parametric derivatives
in the loop expansions.
We first construct 
the lattice EM operators in
the fermionic variables.
Our basic building blocks
are the following:
\begin{align}
  -
  \left.
  \frac{\partial}{\partial \Delta m_n}
  \right \vert_{\rm crit}
  \langle \ldots \rangle_{\rm conn}
  &=
    \Big\langle
    \Big(\frac{1}{2} {\bar{\psi}}_n  \psi_n + 1\Big)
    \ldots
    \Big\rangle_{\rm conn}
    \equiv
    \big\langle
    ( \varepsilon_n+1)
    \ldots
    \big\rangle_{\rm conn},
    \label{eq:m_deriv}
  \\
  \left.
  \frac{\partial}{\partial \kappa_{nm} }
  \right\vert_{\rm crit}
  \langle \ldots \rangle_{\rm conn}
  &=
    \Big\langle
    \Big( {\bar{\psi}}_n P(\vec e_{nm})  \psi_m \Big)
    \ldots
    \Big\rangle_{\rm conn}
    \equiv
    \langle
    E_{nm}
    \ldots
    \rangle_{\rm conn},
    \label{eq:kappa_deriv}
\end{align}
where
the dots represent
other operator insertions
in the path integral
and $\langle \cdot \rangle_{\rm conn}$
the connected part.
The subscript ``${\rm crit}$''
emphasizes that 
we evaluate 
the derivatives at the critical couplings~\eqref{eq:crit_sol_affine}.

By a simple comparison
between the lattice variable $\psi_n$
and the continuum variable $\psi(x)$,
the relative scaling can be
determined. The
$\varepsilon$ operator can then 
be expressed as:
\begin{align}
  \varepsilon(x)
  \simeq
    \frac{2\pi}{a}  \varepsilon_n
  \Leftrightarrow
    -\frac{2\pi}{a}
    \bigg[
    \left.
    \frac{\partial}{\partial \Delta m_n}
    \right \vert_{\rm crit}
    -1
    \bigg].
    \label{eq:eps_ferm_param}
\end{align}
As for the EM tensor, 
we consider
the projected tensor:
\begin{align}
  T_k(x)
  \equiv
  -\frac{1}{2}
  e'_{k \mu}
  e_{k \nu}
  \bar\psi(x) \sigma_\mu \partial_\nu \psi(x),
\end{align}
for which 
we have the lattice 
expression:
\begin{align}
  T_k(x)
  &\simeq
    \frac{2\pi}{a}\frac{1}{\vert \vec \ell'_k \vert}
    \Big[
     E_{n,n+\hat k}
    -
    \frac{1}{2}( \varepsilon_n +  \varepsilon_{n+\hat k})
    \Big]
    \nonumber \\
  &\Leftrightarrow
    \frac{2\pi}{a}\frac{1}{\vert \vec\ell'_k \vert}
    \left.
    \bigg[
    \frac{\partial}{\partial \kappa_{n,n+\hat k}}
    +
    \frac{1}{2}
    \Big(
    \frac{\partial}{\partial \Delta m_n}
    +
    \frac{\partial}{\partial \Delta m_{n+\hat k}}
    \Big)
    -1
    \Big]
    \right \vert_{\rm crit}.
    \label{eq:TM_ferm_param}
\end{align}
The unit vectors $\vec e'_{k}
\equiv
\vec \ell'_k
/
\vert \ell'_k \vert
\equiv
(\cos \alpha'_k,
\sin \alpha'_k)^T
$
encode the angle information
of the $\vec\ell_k'$-lattice.
Up to the EOM,
$T_k(x)$ can be expressed as
a linear combination of 
the holomorphic part $T(z)$ and 
the antiholomorphic part $\tilde T(\bar z)$:
\begin{align}
  T_k(x)
  =
  \frac{1}{2}
  \big[
  \cos(\alpha'_k + \alpha^*_k) T(z)
  -
  \sin(\alpha'_k + \alpha^*_k) \tilde T (\bar z)
  \big].
  \label{eq:mixing_eq}
\end{align}
Thus, the set $T_k(x)$ $(k=1,2,3)$
is sufficient to 
reconstruct all the components
of the EM tensor.

Having obtained the 
lattice operators 
expressed with 
the parametric derivatives,
they can be readily mapped 
to the spin systems
via loop expansions~\eqref{eq:hex_hex}
and \eqref{eq:tri_hex}.
On the hexagonal lattice,
the building blocks~\eqref{eq:m_deriv}
and \eqref{eq:kappa_deriv} 
correspond to:
\begin{align}
  & \varepsilon_n
    =
    \sum_{k}
    \tanh K_{n,n+\hat k} \cosh^2 K_{n,n+\hat k}
    \Big(
    \frac{1}{2}
    \mu_n \mu_m
    +
    \tanh K_{n,n+\hat k}
    \Big)
    +1
    ,
    \label{eq:eps_hex}
  \\
  &E_{n, n+\hat k}
  =
    \frac{\tanh K_{n,n+\hat k} \cosh^2 K_{n,n+\hat k}}{\kappa_{n,n+\hat k}}
    \mu_n  \mu_{n+\hat k},
\end{align}
while on the triangular lattice:
\begin{align}
  & \varepsilon_n
    =
    \sum_{\langle i,j \rangle \textrm{ around } n}
    s_i  s_j
    - 2,
  \label{eq:eps_tri}\\
  &E_{n, n+\hat k}
    =
    -\frac{1}{2}
    \frac{1}{\kappa_{n, n+\hat k}}  s_i s_j.
    \label{eq:E_tri}
\end{align}
In eq.~\eqref{eq:eps_tri},
the sum is over the links
$\langle i,j\rangle$
on the triangular lattice
surrounding 
the hexagonal site $n$,
while in eq.~\eqref{eq:E_tri},
the link $\langle i,j \rangle$
on the triangular lattice
intersects with 
the link $\langle n, n+\hat k \rangle$
on the hexagonal lattice.
The lattice operators
for the CFT, 
$\varepsilon(x)$ and $T_k (x)$,
can then be
constructed by the same linear combinations~\eqref{eq:eps_ferm_param} and \eqref{eq:TM_ferm_param}.

We comment that
both $\varepsilon(x)$ and $T_k(x)$
acquire divergent parts
through loop diagrams
(the divergent part of
$\varepsilon(x)$
comes from the Wilson term).
The divergent parts
need to be subtracted
when we consider disconnected 
components.
This corresponds to 
taking the normal ordering.
For the Ising CFT,
its evaluation can be performed
with the fermion system
without statistical error~\cite{NM2025}.
In the numerical analysis below,
all the lattice operators are
regularized in this way.

\section{Confirming the CFT formulas with Monte Carlo calculations}

In this section,
we calculate various
expectation values 
that involve the EM tensor
with the spin variable
by performing Monte Carlo calculation.
Ensembles are
generated by combining the Wolff algorithm \cite{Wolff:2020oky}
with the local heat-bath algorithm.
The details of the 
calculation 
will be described in Ref.~\cite{NM2025}.
We comment that 
that the signal  
of the EM tensor is noisy
as the 
contributions from 1 (the diverging part), $\varepsilon(x)$, and 
$\partial \varepsilon(x)$ 
(the first descendant of $\varepsilon(x)$)
need to be subtracted
to single out the EM tensor components.
The exact values are
evaluated with the
analytic expressions 
known in the Ising CFT \cite{Eguchi:1986sb,DiFrancesco:1987ez}.
The modulus is set to a 
nontrivial value $\tau=1.2 e^{4i\pi/9}$ below.

\begin{figure}[htb]
  \centering
  \includegraphics[width=4.5cm]{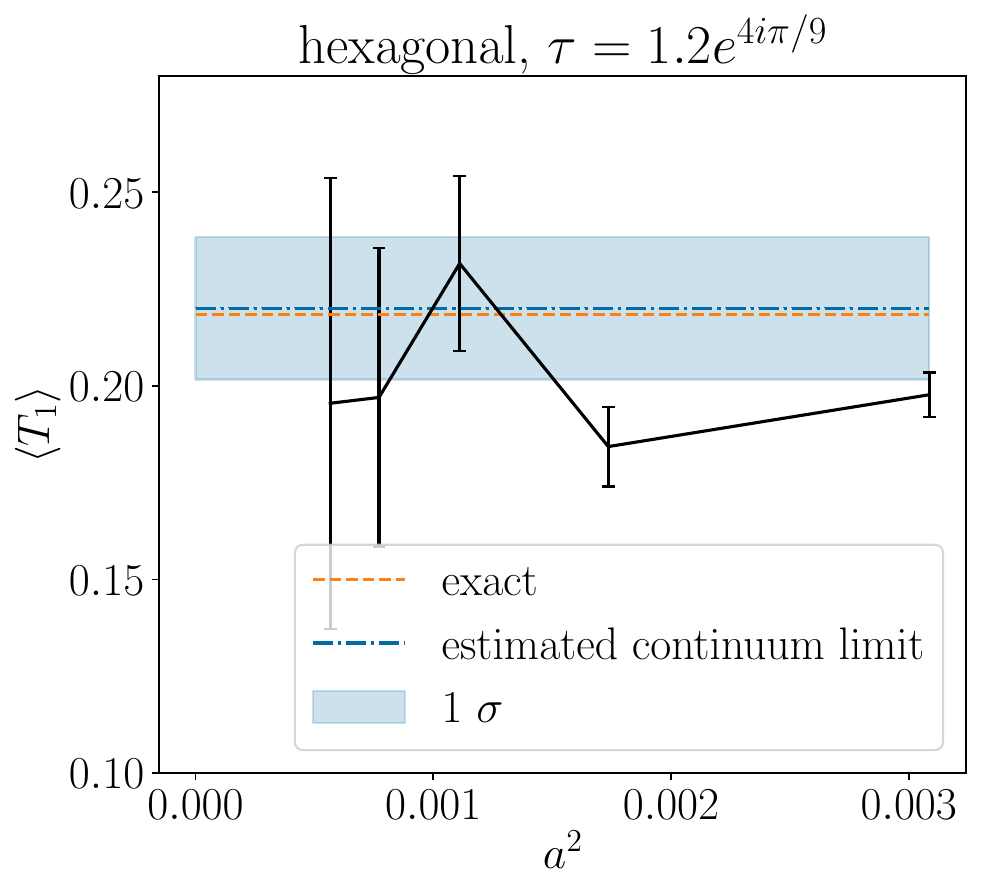}
  \includegraphics[width=4.5cm]{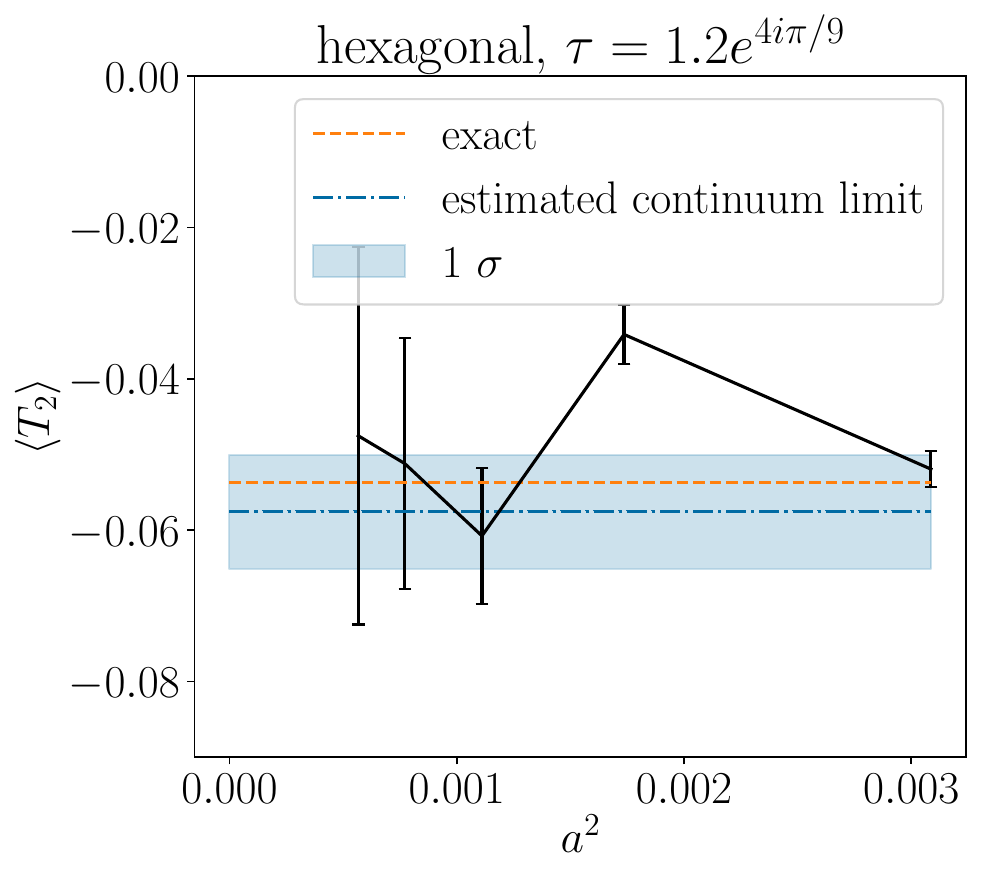}
  \includegraphics[width=4.5cm]{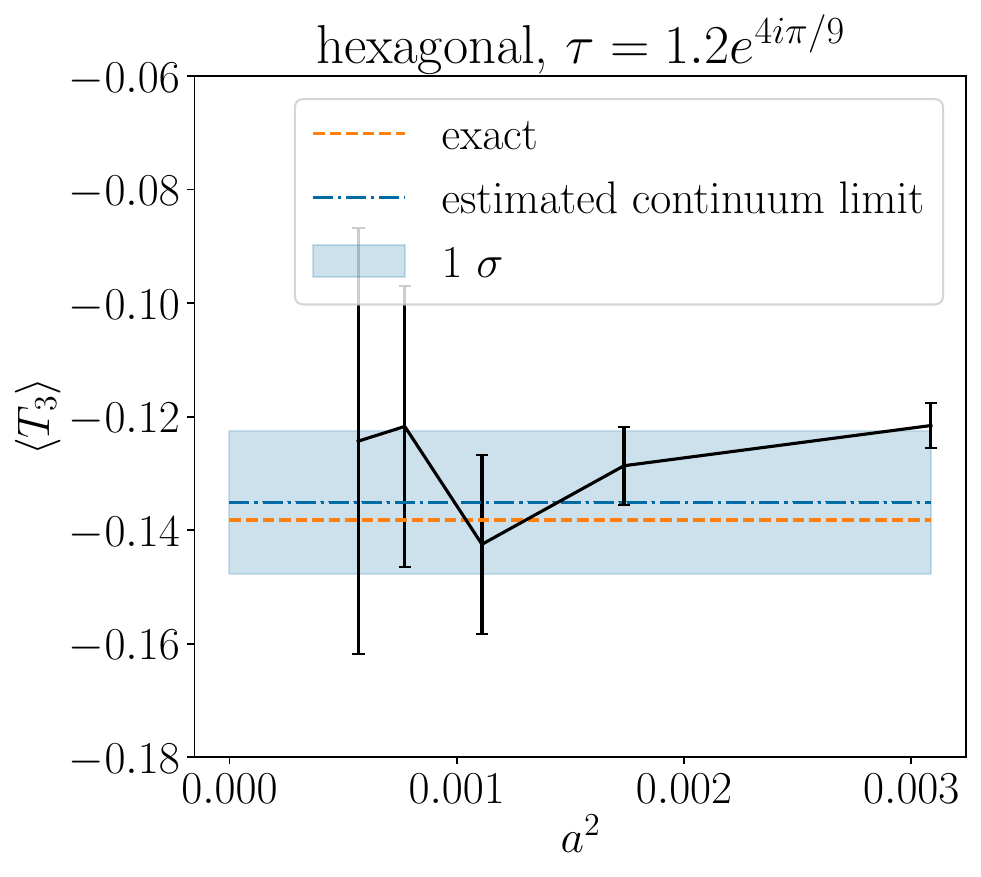}
  \caption{
    Extrapolation of $\langle T_k \rangle$ $(k=1,2,3)$ to the continuum limit
    for $\tau=1.2 e^{4i\pi/9}$
    on the hexagonal lattice for $L=6,8,\cdots,14$.
  }
  \label{fig:T_onept_hex}
\end{figure}

We begin with the one-point function
of the EM tensor.
In Fig.~\ref{fig:T_onept_hex},
we show the results of 
$\langle T_k(x) \rangle$
on the hexagonal lattice.
For completeness,
a constant fit is performed 
for the three points, $L=10,12,14$,
to make a comparison 
with the exact value.
The obtained values are:
$\langle T_1 \rangle \approx 0.215(21)$ (exact: $0.218$), 
$\langle T_2 \rangle \approx -0.0605(88)$ 
(exact: $-0.0537$),
$\langle T_3 \rangle \approx -0.126(15)$ (exact: $-0.138$).
The good agreement shows that
we have a good understanding 
of the lattice operators
including 
the mixing angle 
between $T$ and $\tilde T$
[see eq.~\eqref{eq:mixing_eq}],
the diverging part, and
the normalization.

\begin{figure}[hbt]
  \centering
  \includegraphics[width=4.cm]{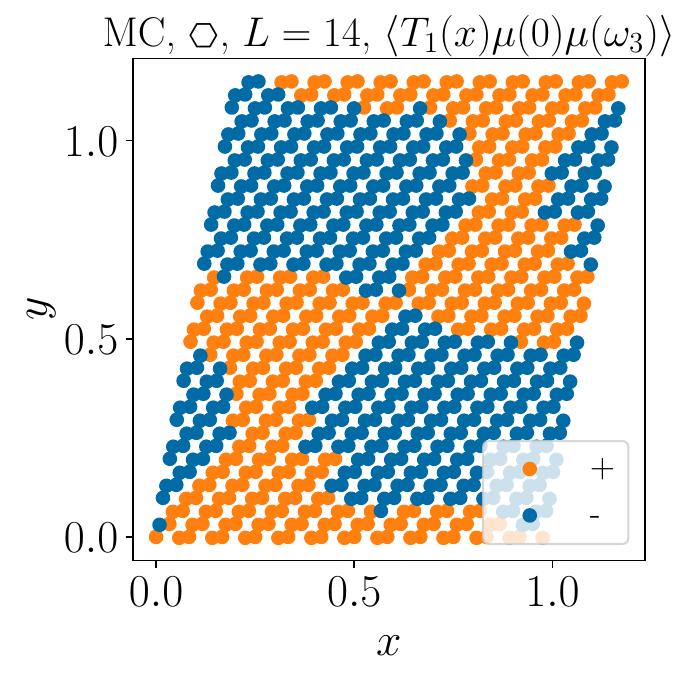}
  \includegraphics[width=4.cm]{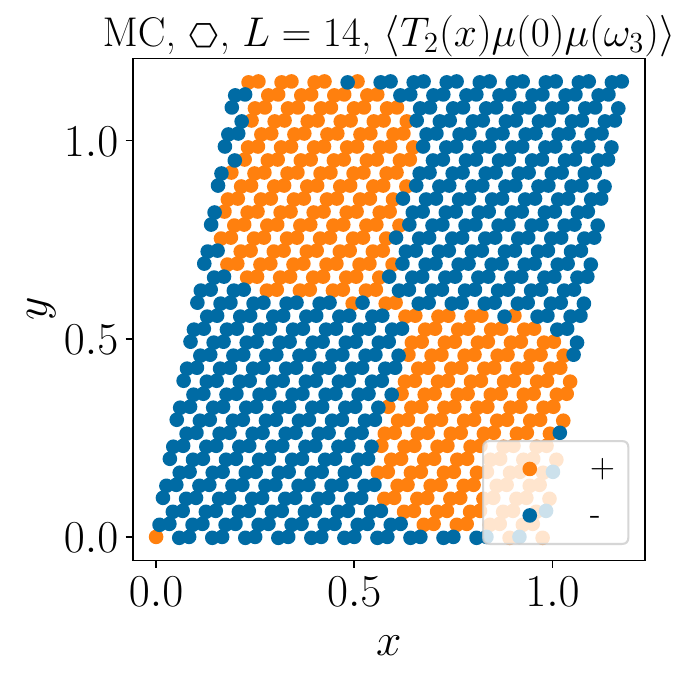}
  \includegraphics[width=4.cm]{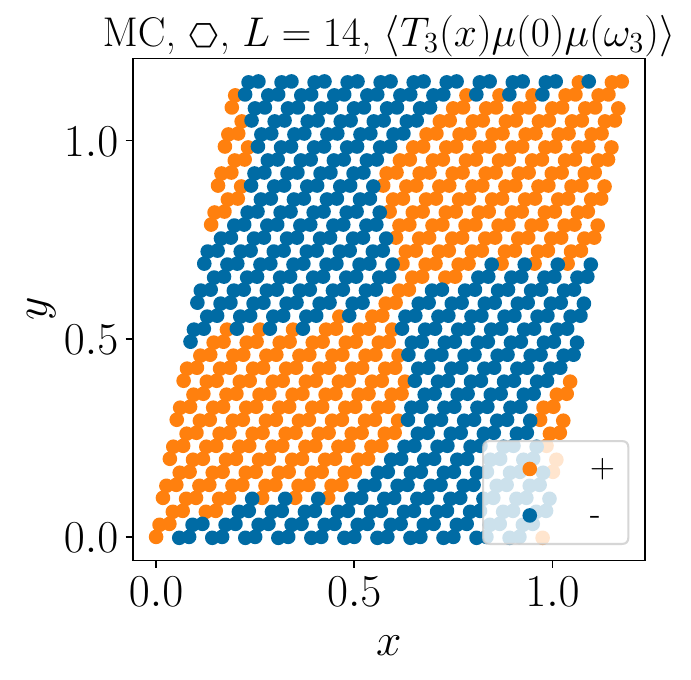}\\
  \includegraphics[width=4.cm]{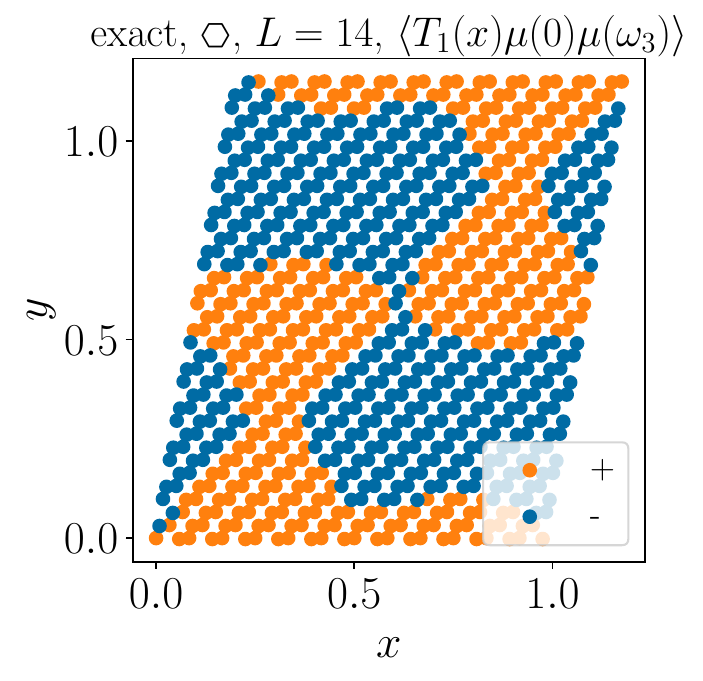}
  \includegraphics[width=4.cm]{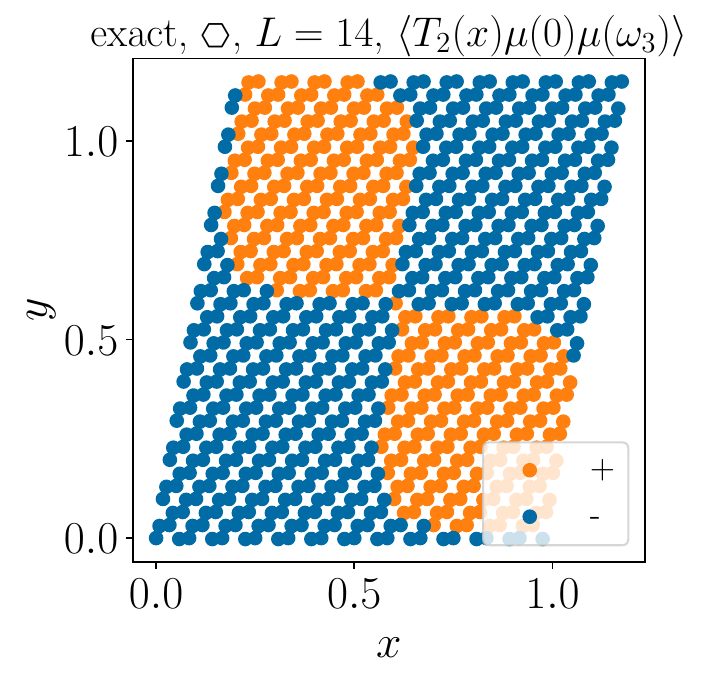}
  \includegraphics[width=4.cm]{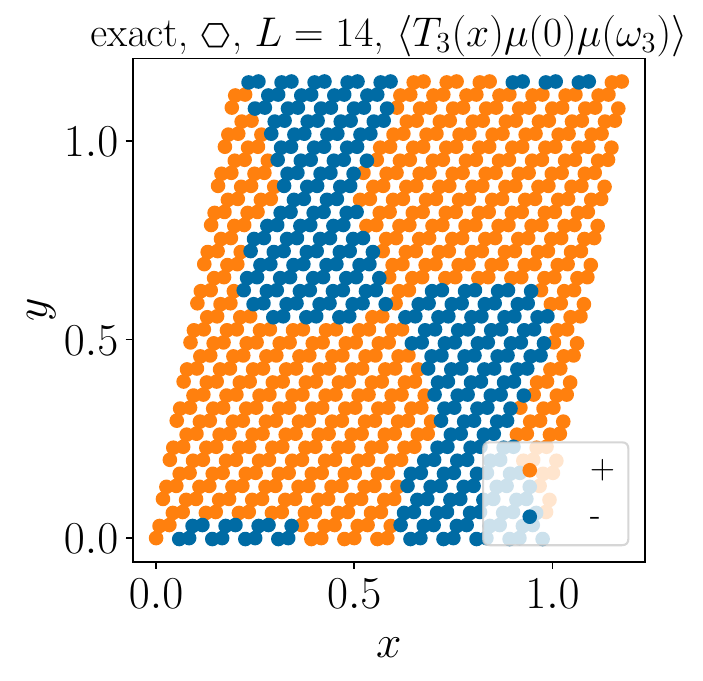}
  \caption{
    The sign patterns of
    $\langle T_k(x) \mu(0) \mu(\omega_3) \rangle$
    on the hexagonal lattice.
    Monte Carlo result (top) and the exact solution (bottom)
    with $\tau = 1.2 e^{4i\pi/9}$, $L=14$,
    and $\omega_3 = (1+\tau)/2$.
    $k=1,2,3$ from left to right.
  }
  \label{fig:Tmumu_sign}
\end{figure}

We next consider the three-point function
with the spin operators.
This is the least noisy correlator
involving the EM tensor.
Recall that 
the conformal Ward identity
for the primary fields 
$\phi_i(z, \bar z)$
with the 
conformal weights $(h_i,\tilde h_i)$
is \cite{Eguchi:1986sb}:
\begin{align}
  &\langle
    T(z) \phi_1(z_1, \bar z_1) \cdots \phi_N(z_N, \bar z_N)
    \rangle
    -
    \langle
    T(z)
    \rangle
    \langle
    \phi_1(z_1, \bar z_1) \cdots \phi_N(z_N, \bar z_N)
    \rangle
    \nonumber\\
  &~~~~~
    =
    \Big\{
    2\pi i \partial_\tau
    +
    \sum_{i=0}^N
    \Big[
    h_i
    \big\{
    \wp(z_i-z) + 2\eta_1
    \big\}
    +
    \big\{
    -\zeta(z_i-z)
    + 2\eta_1 (z_i-z)
    + \pi i
    \big\}
    \partial_{z_i}
    \Big]
    \Big\}\times \nonumber\\
  &~~~~~~~~
    \times
    \langle
    \phi_1(z_1, \bar z_1) \cdots \phi_N(z_N, \bar z_N)
    \rangle,
    \label{eq:ward}
\end{align}
where $\wp(z)$ 
and $\zeta(z)$ are
the Weierstrass functions
and $\eta_1 \equiv \zeta(\omega_1)$ 
($\omega_1\equiv\tau/2$).
We see that 
as $T_k(x)$ approaches
the location of the other 
primary operators, 
it exhibits a 
diverging pole structure 
of order two,
giving rise to a complex global
landscape in its functional form.
For this reason, 
to compare the lattice correlators
with the exact values,
we look into
their sign pattern 
on the entire torus.
Figure~\ref{fig:Tmumu_sign}
shows the result
for $\langle T_k(x) \mu(0) \mu(\omega_3) \rangle$
($\omega_3\equiv(1+\tau)/2$)
on the hexagonal lattice
with $L=14$.
Though we observe a cutoff effect
away from the insertion points $x_i$,
we observe an agreement 
in the global landscape, 
in particular the 
characteristic 
pole structure.

\begin{figure}[htb]
  \centering
  \includegraphics[width=4.cm]{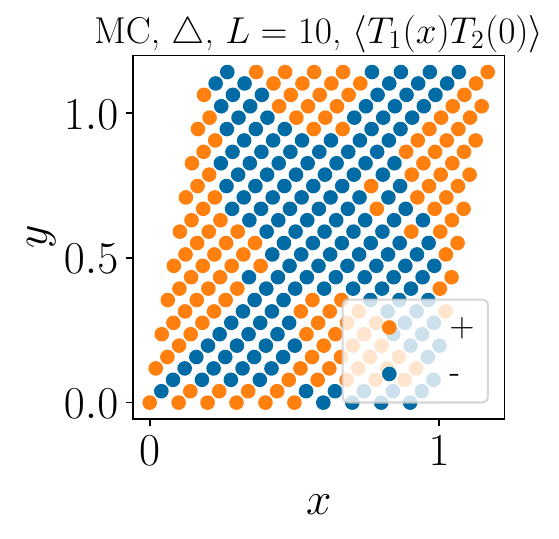}
  \includegraphics[width=4.cm]{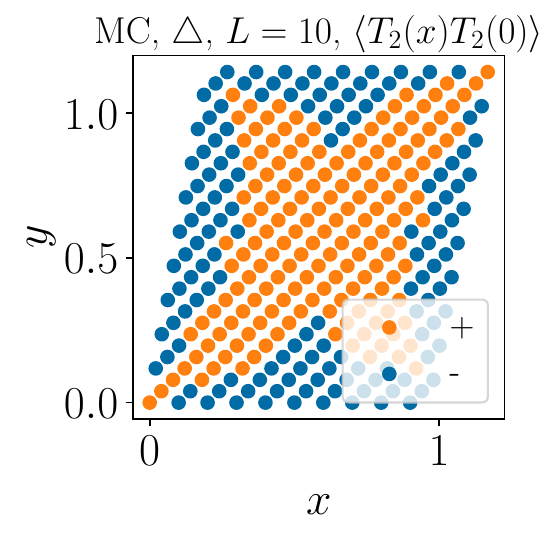}
  \includegraphics[width=4.cm]{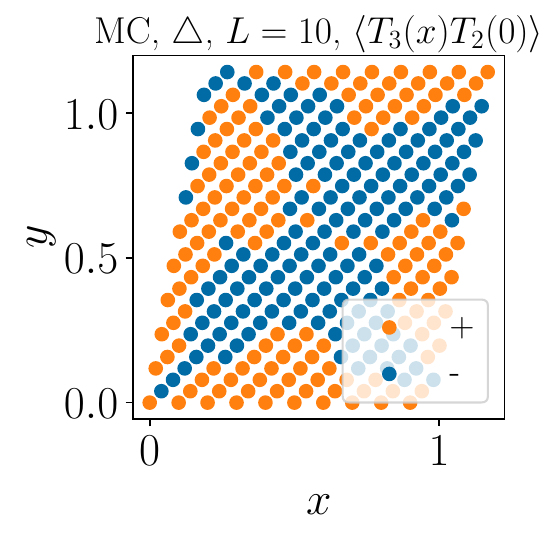}\\
  \includegraphics[width=4.cm]{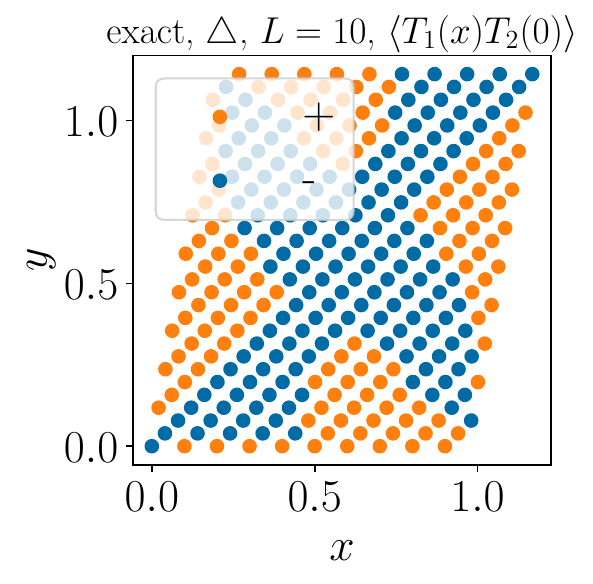}
  \includegraphics[width=4.cm]{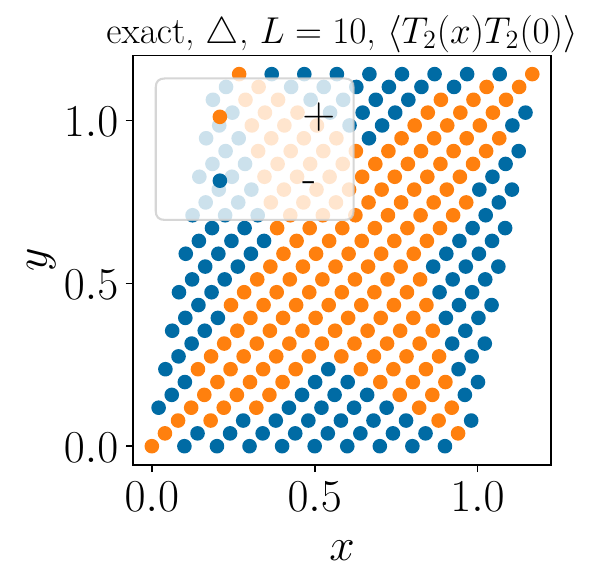}
  \includegraphics[width=4.cm]{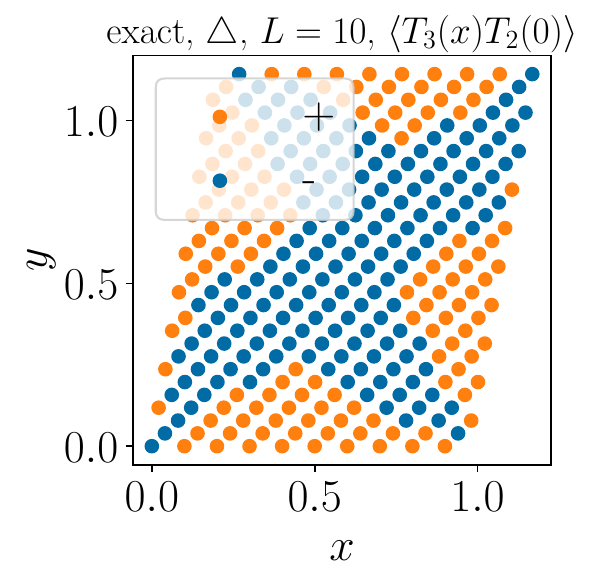}
  \caption{
    The sign patterns of
    $\langle T_k(x) T_2(0) \rangle$
    $(k=1,2,3)$
    plotted for
    the Monte Carlo result (top)
    and the exact solution (bottom)
    with $\tau = 1.2 e^{4i\pi/9}$, $L=6$
    on the hexagonal lattice.
  }
  \label{fig:TT_tri}
\end{figure}

Finally, we calculate the 
$TT$-correlator,
which cannot be re-expressed with
the primary correlators
by the conformal Ward identity.
In Fig.~\ref{fig:TT_tri},
we show the $\langle T_k(x)T_2(0)\rangle$
correlators on the triangular lattice
for $L=10$.
We comment that 
the signal turns out to be less noisy 
on the triangular lattice
than on the hexagonal lattice
for a given $L$
(note, however, that 
the number of lattice sites
is half on the triangular lattice).
We again confirm
the correct landscape of the 
correlator
with statistical noise
at the edge of $\pm$ patterns.

From the above results,
we see that the derived lattice EM tensor correctly sources the
desired CFT operator
both on hexagonal and 
triangular lattices 
under generic affine transformation.

\section{Conclusion}

In this contribution,
we derived the lattice EM tensor
for the Ising CFT with 
spin variables
on triangular and hexagonal lattices
under generic affine transformation.
We numerically confirmed 
that the expression correctly
sources the CFT operator.
It shows our understanding 
is correct on the mixing angle 
between $T$ and $\tilde T$
in relation to the staggered lattice structure,
the identification 
of the divergent part,
and the operator normalization.
Evaluation of
the one-point function 
is of particular importance 
in curved space applications
as it measures 
the trace anomaly 
proportional to the curvature.
We believe that 
this work gives an important 
step towards  
studying the CFT on curved lattices.

\begin{acknowledgments}
The authors thank Hidenori Fukaya, Okuto Morikawa, and Yusuke Namekawa for valuable discussions in LATTICE 2024.
This work is 
partially supported by 
the Scientific Discovery through Advanced Computing (SciDAC) program, 
``Multiscale acceleration: Powering future discoveries in High Energy Physics'' under FOA LAB-2580
funded by the U.S. DOE, Office of Science, and DOE under Award DE-SC0015845.
The computation was performed on the Shared Computing Cluster (SCC) which is administered by Boston University's Research Computing Services \cite{SCC}. 
\end{acknowledgments}

\clearpage

\bibliographystyle{JHEP}
\bibliography{ref}

\providecommand{\href}[2]{#2}\begingroup\raggedright\begin{thebibliography}{10}

\bibitem{Brower:2016moq}
R.C.~Brower, G.~Fleming, A.~Gasbarro, T.~Raben, C.-I.~Tan and E.~Weinberg, \emph{{Quantum Finite Elements for Lattice Field Theory}}, \href{https://doi.org/10.22323/1.251.0296}{\emph{PoS} {\bfseries LATTICE2015} (2016) 296} [\href{https://arxiv.org/abs/1601.01367}{{\ttfamily 1601.01367}}].

\bibitem{Fubini:1972mf}
S.~Fubini, A.J.~Hanson and R.~Jackiw, \emph{{New approach to field theory}}, \href{https://doi.org/10.1103/PhysRevD.7.1732}{\emph{Phys. Rev. D} {\bfseries 7} (1973) 1732}.

\bibitem{Cardy:1985lth}
J.L.~Cardy, \emph{{Universal amplitudes in finite-size scaling: generalisation to arbitrary dimensionality}}, \href{https://doi.org/10.1088/0305-4470/18/13/005}{\emph{J. Phys. A} {\bfseries 18} (1985) 757}.

\bibitem{Brower:2012vg}
R.C.~Brower, G.T.~Fleming and H.~Neuberger, \emph{{Lattice Radial Quantization: 3D Ising}}, \href{https://doi.org/10.1016/j.physletb.2013.03.009}{\emph{Phys. Lett. B} {\bfseries 721} (2013) 299} [\href{https://arxiv.org/abs/1212.6190}{{\ttfamily 1212.6190}}].

\bibitem{Eguchi:1986sb}
T.~Eguchi and H.~Ooguri, \emph{{Conformal and Current Algebras on General Riemann Surface}}, \href{https://doi.org/10.1016/0550-3213(87)90686-9}{\emph{Nucl. Phys. B} {\bfseries 282} (1987) 308}.

\bibitem{DiFrancesco:1987ez}
P.~Di~Francesco, H.~Saleur and J.B.~Zuber, \emph{{Critical Ising Correlation Functions in the Plane and on the Torus}}, \href{https://doi.org/10.1016/0550-3213(87)90202-1}{\emph{Nucl. Phys. B} {\bfseries 290} (1987) 527}.

\bibitem{Kadanoff:1970kz}
L.P.~Kadanoff and H.~Ceva, \emph{{Determination of an opeator algebra for the two-dimensional Ising model}}, \href{https://doi.org/10.1103/PhysRevB.3.3918}{\emph{Phys. Rev. B} {\bfseries 3} (1971) 3918}.

\bibitem{NM2025}
R.C.~Brower, G.T.~Fleming, N.~Matsumoto and R.~Misra{\emph{{,~\it{in preparation}}} }.

\bibitem{PhysRev.65.117}
L.~Onsager, \emph{Crystal statistics. i. a two-dimensional model with an order-disorder transition}, \href{https://doi.org/10.1103/PhysRev.65.117}{\emph{Phys. Rev.} {\bfseries 65} (1944) 117}.

\bibitem{RevModPhys.36.856}
T.D.~Schultz, D.C.~Mattis and E.H.~Lieb, \emph{Two-dimensional ising model as a soluble problem of many fermions}, \href{https://doi.org/10.1103/RevModPhys.36.856}{\emph{Rev. Mod. Phys.} {\bfseries 36} (1964) 856}.

\bibitem{Itzykson:1982ed}
C.~Itzykson, \emph{{Ising Fermions. 1. Two-Dimensions}}, \href{https://doi.org/10.1016/0550-3213(82)90173-0}{\emph{Nucl. Phys. B} {\bfseries 210} (1982) 448}.

\bibitem{Wolff:2020oky}
U.~Wolff, \emph{{Ising model as Wilson-Majorana Fermions}}, \href{https://doi.org/10.1016/j.nuclphysb.2020.115061}{\emph{Nucl. Phys. B} {\bfseries 955} (2020) 115061} [\href{https://arxiv.org/abs/2003.01579}{{\ttfamily 2003.01579}}].

\bibitem{Brower:2022cwv}
R.C.~Brower and E.K.~Owen, \emph{{Ising model on the affine plane}}, \href{https://doi.org/10.1103/PhysRevD.108.014511}{\emph{Phys. Rev. D} {\bfseries 108} (2023) 014511} [\href{https://arxiv.org/abs/2209.15546}{{\ttfamily 2209.15546}}].

\bibitem{10.1063/1.524404}
S.~Samuel, \emph{The use of anticommuting variable integrals in statistical mechanics. i. the computation of partition functions}, \href{https://doi.org/10.1063/1.524404}{\emph{J. Math. Phys} {\bfseries 21} (1980) 2806}.

\bibitem{Seiberg:1986by}
N.~Seiberg and E.~Witten, \emph{{Spin Structures in String Theory}}, \href{https://doi.org/10.1016/0550-3213(86)90297-X}{\emph{Nucl. Phys. B} {\bfseries 276} (1986) 272}.

\bibitem{SCC}
\emph{{Shared Computing Cluster (SCC)}}, {\emph{{}} {\href{https://www.bu.edu/tech/support/research/}{www.bu.edu/tech/support/research/}}}.

\end{thebibliography}\endgroup


\end{document}